\documentclass[useAMS,usenatbib]{mn2e}

\usepackage{graphicx}

\def\aj{AJ}%
\def\araa{ARA\&A}%
\def\apj{ApJ}%
\def\apjl{ApJ}%
\def\apjs{ApJS}%
\def\aap{A\&A}%
\def\cjaa{Chinese J. Astron. Astrophys.}%
\def\icarus{Icarus}%
\def\mnras{MNRAS}%
\def\pasp{PASP}%
\def\nat{Nature}%
\newcommand{\msun}{M_\odot}
\newcommand{\kms}{km\,s$^{-1}$}
\newcommand{\vinf}{v_\infty}
\newcommand{\halfmass}{R_{\rm hm}}
\newcommand{\bmax}{b_{\rm max}}

\title[Multi-planet systems in open clusters]{The dynamical evolution of multi-planet systems in open clusters} \author[Hao, Kouwenhoven \& Spurzem]{
  W. Hao$^{1,2,3}$, 
  M.B.N. Kouwenhoven$^{1}$\thanks{E-mail: kouwenhoven@pku.edu.cn},
  R. Spurzem$^{4,5,1,3}$
  \\
  $^{1}$ Kavli Institute for Astronomy and Astrophysics, Peking University, Yi He Yuan Lu 5, Haidian District, Beijing 100871, P.R.~China\\
  $^{2}$ Department of Astronomy, School of Physics, Peking University, Yi He Yuan Lu 5, Haidian District, Beijing 100871, P.R.~China\\  
  $^{3}$ Max-Planck-Institut f\"{u}r Astrophysik, Karl-Schwarzschild-Str. 1 85741 Garching, Germany \\
  $^{4}$ National Astronomical Observatories of China, Chinese Academy of Sciences, 20A Datun Rd., Chaoyang District, 100012, Beijing, P.R.~China \\
  $^{5}$ Astronomisches Rechen-Institut, Zentrum f\"{u}r Astronomie, Univ. of Heidelberg, M\"{o}nchhof-Strasse 12-14, 69120 Heidelberg, Germany \\
 }

\begin{document}

\date{Accepted ---. Received ---; in original form ---}

\pagerange{---} \pubyear{2013}

\maketitle

\label{firstpage}

\begin{abstract}
  The majority of stars form in star clusters
  and many are thought to have planetary companions. We demonstrate that multi-planet systems are prone to instabilities as a result of frequent stellar encounters in these star clusters much more than single-planet systems. The cumulative effect of close
  and distant encounters on these planetary systems are investigated using Monte Carlo
  scattering experiments. We consider two types of planetary configurations orbiting Sun-like stars: (i) five
  Jupiter-mass planets in the semi-major axis range $1-42$~AU orbiting a Solar mass star, with orbits that
  are initially co-planar, circular, and separated by 10 mutual Hill
  radii, and (ii) the four gas giants of our Solar system. We find that in the equal-mass planet model, $70\%$ of the planets with
  initial semi-major axes $a>40$~AU are either ejected or have collided with the central star or another planet within the lifetime of
  a typical cluster, and that more than $50\%$ of all planets with $a<10$~AU remain bound to
  the system. Planets with short orbital periods are not
  directly affected by encountering stars. However, secular evolution of
  perturbed systems may result in the ejection of the innermost planets or in
  physical collisions of the innermost planets with the host star,
  up to many thousands of years after a stellar encounter. The simulations of the Solar system-like systems indicate that Saturn, Uranus and Neptune are affected by both direct interactions with encountering stars, as well as planet-planet scattering. Jupiter, on the other hand, is almost only affected by direct encounters with neighbouring stars, as its mass is too large to be substantially perturbed by the other three planets. Our results indicate that stellar encounters can account for the apparent scarcity of exoplanets in star clusters,
  not only for those on wide-orbit that are directly affected by stellar encounters,
  but also planets close to the star which can disappear long after a stellar encounter has perturbed the planetary system.  

\end{abstract}


\section{Introduction}

Since the discovery of the first exoplanets by \cite{wolszczan1992} and \cite{mayor1995}, hundreds of new exoplanets have been discovered, and thousands of exoplanet candidates have been identified using the {\em Kepler} observatory \cite[e.g.,][]{batalha2013}. A substantial fraction of the known planets are part of a multi-planet system \citep[e.g.,][]{cumming2008,borucki2011,lissauer2012}. \cite{fabrycky2012} indicate that a large fraction of these exoplanet candidates are genuine, and that the majority of the planets in a multi-planet system orbit their host star in nearly the same plane \citep[see also][]{figueira2012}. 

Most stars, and therefore most planetary systems, are thought to form in star clusters \citep[e.g.,][]{lada2003,clarke2000}, although some may form in relatively isolated regions \cite[e.g.,][]{bressert2010}. A certain fraction of these star clusters may survive for a long time, but many dissolve within 20-50~Myr \citep[see][and references therein]{degrijs2007,portegieszwart2010}. 
Several studies have suggested that even own Solar system may have formed in an open cluster \citep[e.g.,][]{portegieszwart2009}. \cite{dukes2012}, on the other hand, argue that current observations do not rule out that the Solar system may have formed in a massive, but short-lived,  star cluster. We refer to \cite{adams2010} for an extensive review on the birth environment of the Solar System. 
In such dense stellar environments, dynamical encounters between stars are frequent and can perturb the orbit of binary stars and planetary system significantly. 

As star clusters are the likely birth places of the majority of the planet-hosting stars in the Solar neighbourhood, numerous exoplanet surveys have been carried out in open clusters and globular clusters, but their results have been disappointing \citep[e.g.,][and references therein]{zhou2012}. Very few or no close-in planets with orbital period range of $1-16$ days have been detected in 47~Tucanae and $\omega$~Centauri \citep{weldrake2005,weldrake2008}, the Hyades \citep{guenther2005}, NGC\,6397 \citep{nascimbeni2012}, NGC\,6791 \citep{montalto2007}, M37 \citep{hartman2009}, NGC\,6791 \citep{mochejska2005}, NGC\,2158 \citep{mochejska2006}, NGC\,7789 \citep{bramich2006},  NGC\,1245 \citep{burke2006}, and  NGC\,7068  \citep{rosvick2006}.   
However, a number of planets have been found in open clusters: \cite{lovis2007} and \cite{sato2007} both report a planet in a wide orbit around a giant star. In addition, \cite{quinn2012} report the discovery of two hot Jupiters in Praesepe, making these the first hot Jupiters to be detected orbiting a main sequence star in a star cluster. Finally, protoplanetary disks have been detected in star clusters \citep[e.g.,][and references therein]{hillenbrand2005}, hinting at planet formation. These observations demonstrate that planets can form in star clusters, but it leaves the question why so few have been discovered.

The paucity of planets in star clusters may partially be explained by the suppression of planet formation in star clusters. Jupiter-mass planets are less likely to form in the neighbourhood of massive stars in star clusters due to the evaporation of circumstellar disks by photo-evaporation \citep{armitage2000,holden2011}. \cite{olczak2006,olczak2008}, \cite{forgan2009} and \cite{lestrade2011} showed that stellar encounters reduce planet formation due to disk destruction, although \cite{thies2005,thies2010} find that encounters between circumstellar disks and other stars may actually enhance planet formation.

After the formation process, a newly-formed planetary system and its cometary population remain prone to disruption by encountering stars during the lifetime of the star cluster \citep[e.g.,][]{adams2001,adams2006a,adams2006b,brasser2012}. The importance of the encounters is primarily determined by the local stellar density, the collisional cross section of the planetary system, and the timescale at which the system is exposed to external perturbations. \cite{bonnell2001} and \cite{smith2001} show that planetary systems similar to our own Solar Systems are likely to survive in open clusters, but not in globular clusters.
\cite{fregeau2006} study star-planet systems in star cluster environments and derive dynamical cross-sections and the location of the hard-soft boundary for star-planet systems, showing that these systems are more prone to disruption than previously thought.
Flybys can cause immediate ejections, but can also trigger instabilities in multi-planet systems that can lead to the ejection of one or more planets millions of years after the encounter occurred \citep{malmberg2007,malmberg2011}. A systematic study on the evolution of single-planet systems in globular clusters was carried out by \cite{spurzem2009}. They find that close-in planets are difficult to disrupt, but their eccentricities are excited by weak encounters, which may ultimately result in their migration as a result of tidal evolution. 

Given that most star clusters may initially have a cool virial state and a fractal structure, close encounters may be more frequent during the early phase of relaxation \citep{woolfson2004,parker2012}. A highly inclined companion could excite planets to high eccentricity orbits through the Kozai mechanism \citep{kozai1962} as a result of secular interaction \citep{wu2003}. Studies on the secular instability of several isolated multi-planet systems caused by angular momentum deficit (AMD) and mean motion resonance show that these non-circular and non-co-planar orbits generated by the perturbation with random resonances would be unstable \citep{laskar2009,laskar2000}. 

The evolution of compact multi-planet systems in star clusters, however, remains poorly understood, although previous studies have already indicated that the effect of planet-planet interactions may be at least as important as the effect of stellar flybys \citep[][]{adams2006a,adams2006b,malmberg2011,boley2012}.
It is difficult to model the evolution of these in star clusters using direct $N$-body simulations, because of the enormous range of temporal and spatial scales between individual planetary systems and their host star clusters. In addition, the number of weak stellar encounters is large and numerical errors in planetary systems are cumulative \citep[e.g.,][]{aarseth2003}. Previous studies have therefore often focused on single-planet systems, which can be modelled as perturbed two-body systems, or on planetary systems in low-mass, short-lived star clusters.

In this article we present Monte Carlo scattering experiments to study the evolution of multi-planet systems in dense stellar environments, modelling all encounters within $1000$~AU during the time a  typical planetary system spends in a dense open cluster environment. We find that the effect of flybys is important for the evolution of a multi-planet system. Interestingly, we demonstrate that the evolution of short-period planets is substantially affected by the presence of outer planets.
The article is organised as follows: we describe the methods and assumptions in
\S~\ref{section:method}. The results and analysis are presented in
\S~\ref{section:results}, and finally we draw the conclusions in \S~\ref{section:conclusions} and
discuss our results in \S~\ref{section:discussion}.


\section{Method and initial conditions} \label{section:method}

Under some circumstances it is possible to
run full $N$-body simulations of dense star clusters with
planetary systems. For single-planet systems, this is
relatively easy, since two-body systems can be
regularised. \cite{spurzem2009}, for example, used this method for a comprehensive
study on the dynamical evolution of single-planet systems in populous
star clusters. The situation is substantially more complex for multi-planet systems. Individual planetary motions require integration using time steps which are many orders of
magnitude smaller than those used for the stellar dynamics. In addition, the integration errors in planetary systems are cumulative. As the
secular behaviour of a multi-planet system is of utmost importance to
its dynamical fate, it is currently impractical to use full
$N$-body simulations.  Note that several attempts have been made to carry
out such simulations with relatively low-mass star clusters and on
shorter timescales \citep[e.g.,][]{malmberg2011}. To study the
evolution of multi-planet systems in dense clusters on long
timescales, however, we need to resort to alternative techniques.

We use a Monte-Carlo scattering approach to simulate the environment
in which most of the planetary systems are born. We model the
evolution of the newly generated planetary systems at high
accuracy and attempt to reproduce their environments more
realistically than in previous studies. Instead of carrying out
time-consuming, and for our purpose inaccurate, direct $N$-body
simulations of star clusters with planetary systems, we focus only on
individual planetary systems and their encountering stars, while
ignoring the more or less homogeneous effect of other surrounding
stars in the cluster.

We first calculate the encounter rate for
a typical star in an open cluster environment. Based on
the total number of stars that approach the planetary
system within a distance of ~1000~AU during its lifetime in the star
cluster, we let the encountering stars approach the planetary system
one by one at a Poissonian time interval, derived from the average
encounter frequency.  

\subsection{Modelling the encountering stars}  \label{section:encountermodeling}

We consider the evolution of planetary systems in open
clusters similar to the Orion Nebula Cluster (ONC), which has roughly
$N=2\,800$ stars with a one dimensional velocity dispersion $\sigma =2.34$~\kms\  \citep{hillenbrand1998} and a half-mass
radius of $\halfmass=0.5$~pc. Within the half-mass radius, the average
distance between two stars is approximately $\halfmass N^{-1/3}$. We assume that the
star cluster is in virial equilibrium and ignore the effects of mass
segregation and evaporation.  The lifetime of an open cluster depends on
the properties of the star cluster and its environment. Open clusters dissolve within
$10^7-10^9$ years \citep[e.g.,][]{degrijs2007} and we therefore adopt a
total integration time of 100~Myr.

The properties of the encountering stars are modelled by drawing (i)
the relative velocity at infinity $\vinf$, (ii) the impact parameter $b$, (iii)
the mass $M_e$ of the encountering star, and (iv) the encounter rate
$\Gamma$ from the distributions that are representative for open
clusters. For simplicity, we assume in this study that these
four parameter distributions are mutually independent, and independent
of time. All encountering stars are initialised to approach from
random directions.  Our choices for these four parameters, as well as
several other environmental quantities are summarised in
Table~\ref{table:encounters}.

The masses $M_e$ of the encountering stars are drawn from the
\cite{chabrier2003} normal mass distribution,
\begin{equation} \label{equation:chabrier}
  f(\log M_e) \propto \frac{1}{\sigma_\mu\sqrt{2\pi}} 
\exp \left( -\frac{ (\log M_e-\mu)^2 }{2\sigma_\mu^2} \right) \ ,  
\end{equation}
where $\mu = \log(0.2)$ and $\sigma_\mu = 0.55$ in the mass range
$0.08-5\msun$. Although more massive
stars can cause substantial damage to planetary systems during their
encounters, we ignore them since they are relatively rare and
short-lived. As mentioned above, we assume that the mass of the encountering stars is
uncorrelated with their impact parameter, their velocity at infinity, and time. In other words, we ignore the effects of
mass segregation and stellar evolution.

The encounter velocities are derived from the velocity dispersion of
the star cluster. For the \cite{plummer1911} model,
its relation to the total mass $M_{\rm cluster}$ and half-mass radius $\halfmass$ of the cluster is given by
\begin{equation}
  \sigma = \sqrt{\frac{GM_{\rm cluster}}{\eta\halfmass}} ,
\end{equation}
where $\eta\approx 9.75$, and $G$ is the gravitational
constant \citep[e.g.,][]{spitzer1987, heggiehut}. In our simulations
we are interested in the distribution of velocity differences
$\vinf$ between two stars, which can be obtained from the
Maxwell-Boltzmann distribution,
\begin{equation} \label{equation:maxwellian}
  f(\vinf) = \frac{\vinf^2 }{2\sqrt{\pi}\sigma^3}\exp
\left( -\frac{\vinf^2}{4\sigma^2} \right) \ ,
\end{equation}
where the mean velocity difference between two stars is $\langle \vinf \rangle =2\sigma\sqrt{2\pi^{-1}} = 3.74$~\kms, and $\sigma$
is the one-dimensional velocity dispersion.  As all planets have
masses much smaller than either of the two stars involved, the stars follow
near-hyperbolic trajectories. The relative velocity $v(r)$ between the
two stars as a function of their separation $r(t)$ can therefore be approximated with
\begin{equation}
  v(r) = \frac{2GM_T}{r(t)} + \vinf  \ ,
\end{equation}
where $M_T=M_c+M_e$ denotes the combined mass the bodies involved. The
hyperbolic orbit can be described using its impact parameter $b$ and
its velocity at infinity $\vinf$. The impact parameter $b$ is drawn
from the distribution
\begin{equation} \label{equation:impactparameter}
  f_b(b) = \frac{2b}{\bmax^2} \ .
\end{equation}
We only consider encounters with impact parameter $0 < b \leq \bmax$,
where $\bmax$ is the maximum impact parameter, and we ignore the
effect of stars with an impact parameter larger than $\bmax$.
\cite{adams2001,adams2006b} have shown that the most
important parameter that determines the effect of a stellar encounter
on a planetary system is their distance of closest approach $p$. Its relation to the impact parameter $b$ can be expressed as
\begin{equation}
 b = p\sqrt{1+\frac{2GM_T}{p\vinf^2}}\ .
\end{equation}
Typical encounters have $\langle M_T \rangle \approx 1.55\msun$ and
$\langle \vinf \rangle = 3.74$~\kms. For these interactions, $b>1094$~AU corresponds to a 
distance of closest approach of $p > 1000$~AU.  In our simulations we
ignore the weak perturbations induced by stars with $p>1000$~AU.

Now that we have drawn $\vinf$, $b$, and $M_e$ from their respective
distributions, we can calculate the hyperbolic semi-major axis
\begin{equation}
  a = \frac{GM_T}{\vinf^2} = \frac{b}{\sqrt{e^2-1}} = \frac{p}{e-1}
\end{equation}
and the hyperbolic eccentricity
\begin{equation}
  e = \sqrt{1 + \left( \frac{b\vinf^2}{GM_T}  \right)^2 } \ ,
\end{equation}
where $p$ is the distance of closest approach. Finally, the velocity at
periastron $v_p$ is given by
\begin{equation}
  v_p^2 = \frac{GM_T}{a} \left( \frac{e+1}{e-1} \right) \ .
\end{equation}

The encounter rate $\Gamma$ is calculated using $\Gamma = n\sigma A$,
where $n$ is the local stellar density, $\sigma$ the velocity
dispersion and $A$ the cross section of the encounter, which can be
approximated with
\begin{equation}
  A = \pi b^2 = \pi p^2 \left( 1 + \frac{v_p^2}{\vinf^2} \right) 
  = \pi p^2 \left( 1+ \frac{e+1}{e-1} \right) \ ,
\end{equation}
The time $\tau$
between two subsequent interactions is drawn from a Poisson
distribution
\begin{equation} \label{equation:poisson}
  P(N(t)=r)=\frac{(t\,\Gamma^{-1})^r e^{-t\,\Gamma^{-1}}}{r!}
\end{equation}
where $\langle \tau \rangle = \Gamma^{-1}$ is the average time between
two subsequent encounters. For the star cluster listed in
Table~\ref{table:encounters} we obtain $\langle \tau \rangle
\approx 1.7\times 10^6$~year, corresponding to a total number of 58
stars that approach the planetary system within $p=1000$~AU.

\begin{table}
  \begin{tabular}{ll}
    \hline
    Quantity                        & Value                   \\
    \hline 
    Number of stars $N$             & $2800$             \\
    Initial mass function           & Chabrier $0.08-5\msun$  \\
    Half-mass radius $\halfmass$    & 0.5~pc                  \\
    Velocity dispersion $\sigma$    & 2.34~\kms               \\
    Closest approach $p$ (AU)		& $0-1000~AU$		\\
    Impact parameter $b$ (AU)       & $f_b(b) \propto b$; $\bmax =$ 1094~AU      \\
    Mean velocity difference $\vinf$		& 3.74~\kms		\\
    Average encounter interval      & $\langle \tau \rangle = 1.7$~Myr; Poissonian \\
    Total integration time          & $T=100$~Myr                 \\
    Encounter number 		& 58                      \\
    Impact direction                & Random                  \\
    \hline
  \end{tabular}
  \caption{Initial properties of the example open cluster and stellar encounters. 
 \label{table:encounters} }
\end{table}


\subsection{Initial conditions for planetary systems} \label{section:initialconditions}

We model two sets of planetary systems, which we refer to as model~1 and model~2, respectively. The initial conditions for the planetary systems are summarised in
Table~\ref{table:initialplanets}. We adopt these two sets of planetary configurations in order to obtain a better understanding of the effect of different configurations of multi-planet systems.

For model~1 we follow the EMS
(equal-mass, equal-relative-separation systems) prescriptions of
\cite{zhou2007} and we adopt a central star mass of
$1\msun$. Planet formation simulations have indicated that when the mass of a gas giant reaches around a Jupiter mass ($1 M_J$), its oligarchic growth terminates and the planet obtains its isolation mass \citep{kokubo2002, ida2004}. Therefore, we assign all planets a  mass of $1 M_J$. These simulations also indicate that most planetary embryos are  initially separated by $10-12$ mutual Hill radii ($R_H$) after they have depleted the planetesimal populations in their neighbourhoods. We therefore assign our planets orbits which are initially separated by $k=10 R_H$. The mutual Hill radius for two
neighbouring orbits with semi-major axes $a_1$ and $a_2$ is defined as
\begin{equation}
  R_H = \frac{a_1+a_2}{2}\left( \frac{M_1+M_2}{3M_c} \right)^{1/3}
\end{equation}
where $M_1$ and $M_2$ are the masses of the two planets and $M_c$ is
the mass of the central star.
 
For the innermost planet we adopt a semi-major axis of $a_{\rm min}=1$~AU,
which is within the ice line for solar-type stars, and accounts for a
modest amount of migration. The semi-major axis of the subsequent
planets is determined by the value of $k$. In our case we choose $N_p=5$
planets, resulting in semi-major axes for the planets of $a=1$, 2.6,
6.5, 16.6, and 42.3~AU, where the outermost semi-major axis reaches the
limit at which planetary systems are able to form planets through the
core-accretion process. All planets are assigned zero initial
eccentricities ($e=0$) and inclinations ($i=0^\circ$), and the initial
orbital phase of each planet is drawn randomly from
$\phi \in [0^\circ,360^\circ]$.  

For model~2 we adopt a Solar-mass star with four gas giants: Jupiter, Saturn, Uranus, and Neptune. Each of these planets is assigned their mass and present-day orbital elements identical to their counterparts in our Solar System. Note that the encountering stars come from all
directions, so the inclinations of all orbits will gradually increase over time.

\begin{table}
  \begin{tabular}{lll}
    \hline
    Model & Quantity                 & Value \\
    \hline
    1,2 & Central star mass       & $M_c = 1\msun$ \\
    \hline
    1   & Number of planets       & $N_p=5$\\
    1   & Planet mass             & $M_p = 1~M_J$ \\
    1   & Minimum semi-major axis & $a_{\rm min}= 1$~AU\\
    1   & Mutual separation       & $k=10$\\
    1   & Eccentricity            & $e=0$\\
    1   & Inclination             & $i=0^\circ$\\
    1   & Initial orbital phase   & $0^\circ \leq \phi < 360^\circ$ \\
        \hline
    2   & Number of planets       & $N_p=4$ \\
    2   & Orbital parameters      & Jupiter, Saturn, \\
        &                         & Uranus, and Neptune \\
    \hline
  \end{tabular}
  \caption{Initial conditions for the planetary systems used in our simulations, for model~1 and model~2, respectively. Note that model~2 represents our Sun with the four gas giants in our Solar system. \label{table:initialplanets} }
\end{table}

\subsection{Simulations} \label{section:simulations}

The simulations are carried out by combining the strengths of two integration
packages. During an encounter with the planetary system, the package
CHAIN is used, while the MERCURY package is used to model the
long-term secular evolution of the system in between two subsequent
encounters.
The CHAIN package \citep{mikkola1993,aarseth2003} uses chain
regularisation and is particularly useful for the treatment of
energetic multi-body encounters. The integration is carried out
quickly, and energy is conserved at a level of $|\Delta E/E| < 10^{-13}$.
Although CHAIN can very accurately model short-term energetic encounters in
small-$N$ systems, it is not optimised for long-term evolution of
isolated planetary systems. To this end, we use the MERCURY
\citep{chambers1999} package to model the evolution of the system in
between two encounters. MERCURY is a hybrid integrator that allows
symplectic integration of planetary systems over longer timescales, and can accurately deal with planet-planet scattering. The simulations typically
conserve energy at the $|\Delta E/E| = 10^{-6}-10^{-8}$ level.
The switch from the CHAIN package to the MERCURY package and vice-versa is
made when the encountering star is sufficiently far from
the planetary system ($r>1000$~AU), such that its effect on the outermost planet in the system is negligible.  

Star-planet and planet-planet collisions are modelled by merging two
objects that enter each other's Roche lobe. In the case of a physical collision, they are replaced by a new particle with the combined mass of the
two objects and with a position and velocity of their centre-of-mass.

Throughout our simulations we continuously check for escaping planets,
which are identified as follows. All planets at a distance $r > R_{\rm
  lim}$ from their host star are removed from the system, where $R_{\rm lim}$ is the average separation between the members of the star cluster. Planets at a distance $500~{\rm AU} < r < R_{\rm lim}$ are removed when (i) their binding energy $E$
with respect to the central star is positive, (ii) a distance to the
central star larger than 20\,000~AU, and (iii) their velocity
vector points away from the central star ($\vec{r}\cdot
\vec{v}>0$).  After escaper removal we calculate the velocity at
infinity for each escaper correcting for the potential
energy at the moment of removal.

To improve statistics we carry out simulations for an ensemble of a hundred
planetary systems. Each simulation starts out with an identical planetary system with randomly drawn initial orbital phases. The properties of the encountering stars and the time scales between two subsequent encounters drawn randomly from the distributions mentioned above.


\section{Results} \label{section:results}

\begin{figure}
  \centering
  \includegraphics[width=0.45\textwidth,height=!]{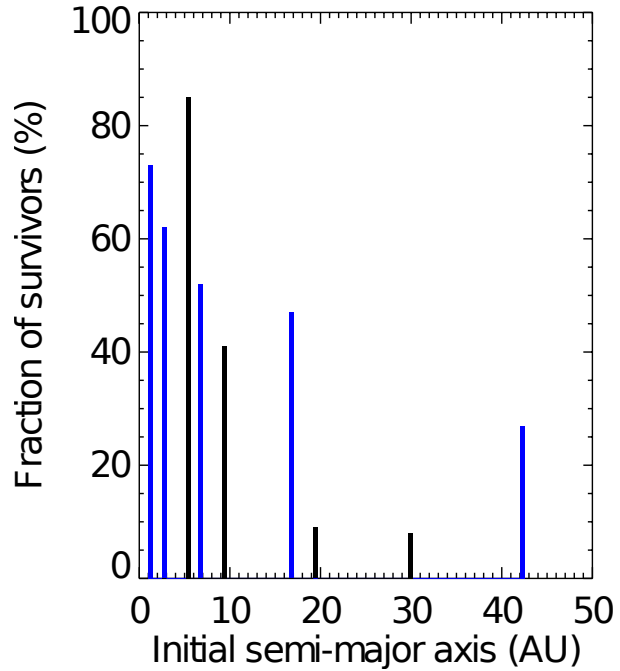}
  \caption{Fraction of survived planets as a function of their initial semi-major axis, for model~1 (blue) and model~2 (black). Inner planets have a higher probability of surviving after the stellar encounters, while outer planets can more easily escape the system.
    \label{figure:bound_versus_position} }
\end{figure}

\subsection{Final planet configuration}

A planetary system experiences many encounters during the time which it spends in the star cluster. The cumulative effect of these encounters may result in the ejection of one or more of the planets, star-planet and planet-planet collisions, or a reconfiguration of the planetary system (see \S~\ref{section:simulations} for details on the procedures involved). 

After simulating a hundred runs of target planetary systems within their lifetime in the open cluster ($\sim 100$~Myr), we obtain statistical results by combining the outcomes. Figure~\ref{figure:bound_versus_position} shows the fraction of planets that is still bound to their host star, as a function of their initial semi-major axis. As expected, planets with the smallest initial semi-major axis are more likely to survive. For model~1 a clear decline in the survival chances is seen with increasing initial semi-major axis. More than $50\%$ of the planets with $a<10$~AU have survived, while less than $30\%$ of planets with $a>40$~AU are bound to the system at the end of the simulations. The highest retention rate is found for planets with $a=1$~AU, with a survival probability of more than $70\%$. For model~2 we see a similar, but steeper trend. A relatively large fraction of the Jupiters survive, while most of the outer planets escape the system. This steeper trend is a result of the mass spectrum of the planets in model~2: it is more difficult to eject the high-mass Jupiters, while it is easier to remove Uranus and Neptune from the system.

\begin{figure}
  \centering
  \includegraphics[width=0.45\textwidth,height=!]{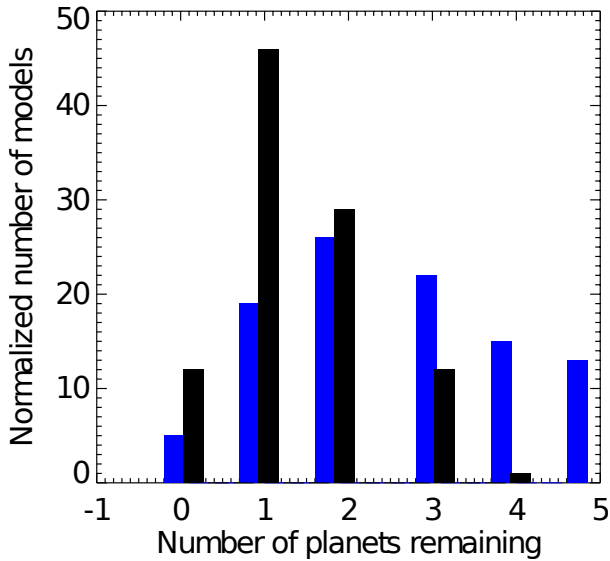}
  \caption{The distribution of number of planets still bound to the host star at the end of the simulation, for model~1 (blue) and model~2 (black). At the end of the simulations, only $13\%$ and $1\%$ of the host stars have all their five planets in orbit, while $5\%$ and $12\%$ of the host stars have lost all their planets, for model~1 and model~2, respectively.
    \label{figure:total_bound} }
\end{figure}

The fraction of planets that is still bound after many encounters within the lifetime of its parent cluster is undisputedly the most important parameter in assessing the efficiency of the combined effect of external perturbations and internal secular evolution in planetary systems. We therefore examine the statistics of planet retention for all our data and plot the results in Figure~\ref{figure:total_bound}. In the equal-mass system (model~1), 5\% of the systems have lost all their planets, while for 13\% of the systems all planets remain bound despite the repeated stellar encounters. The peak in the distribution gradually shifts to lower numbers over time. In model~2, on the other hand, 12\% of the stars have lost all their planets, while only 1\% of the systems retain all their planets. It is worth noting that almost half of the systems only has one planet (Jupiter) left at the end of the simulations. As explained above, this is caused by the unequal planetary masses in this configuration.

\begin{figure}
  \centering
  \includegraphics[width=0.45\textwidth,height=!]{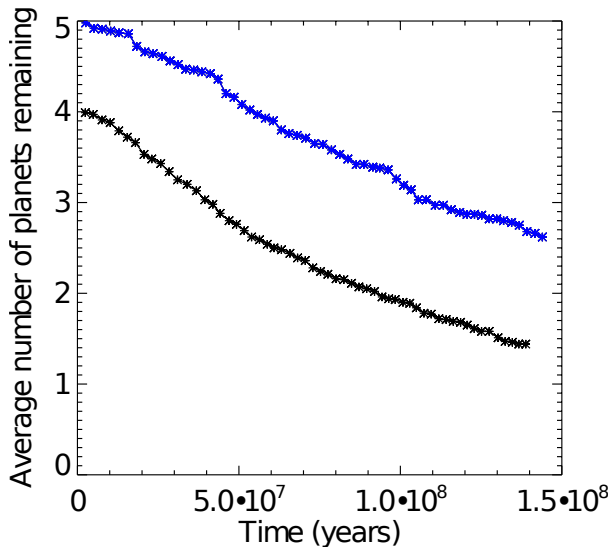}
  \caption{The average number of planets bound to the central star as a function of time, for model~1 (blue) and model~2 (black). 
  \label{figure:bound_time} }
\end{figure}

\begin{figure}
  \centering
  \begin{tabular}{c}
  \includegraphics[width=0.45\textwidth,height=!]{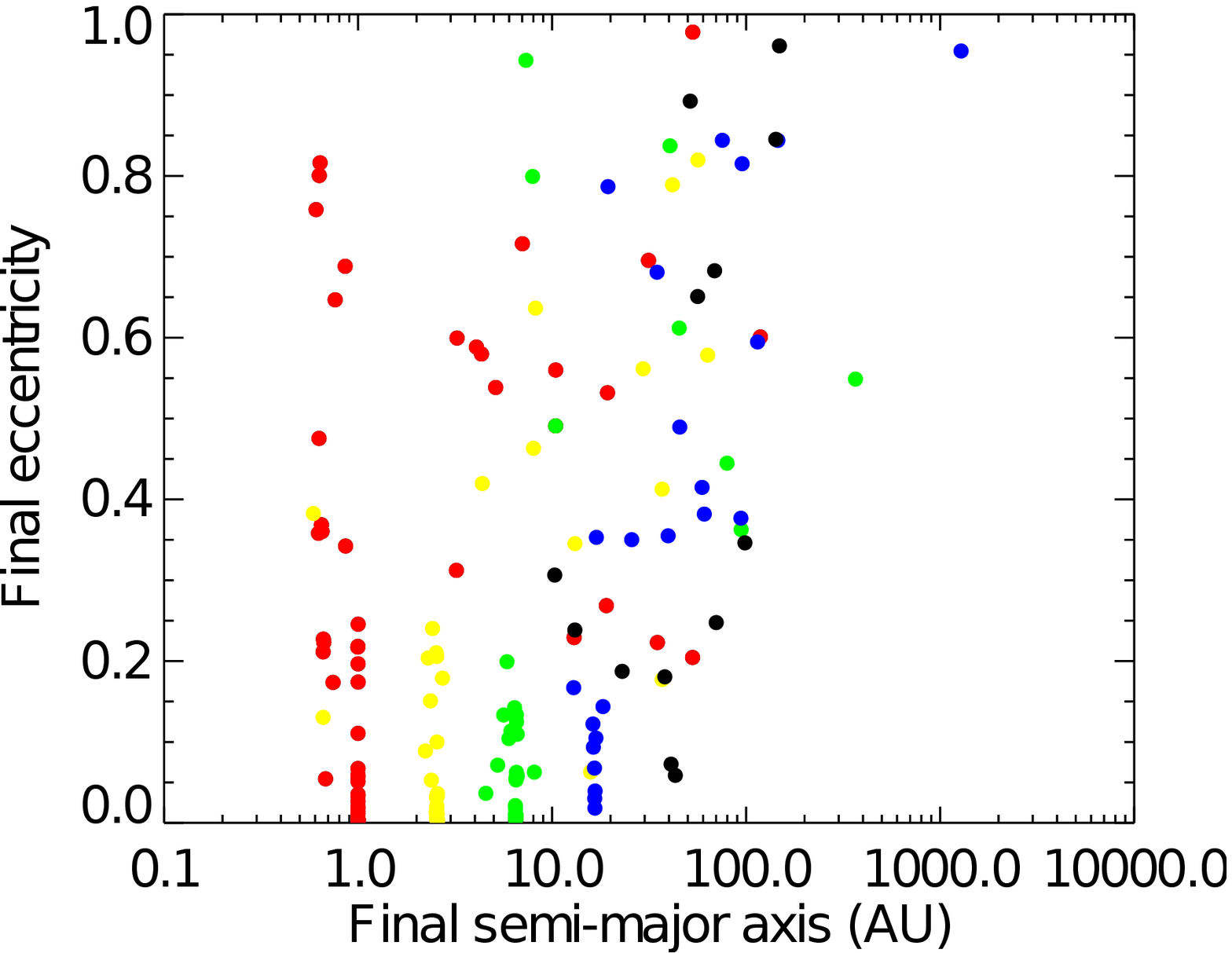} \\
  \includegraphics[width=0.45\textwidth,height=!]{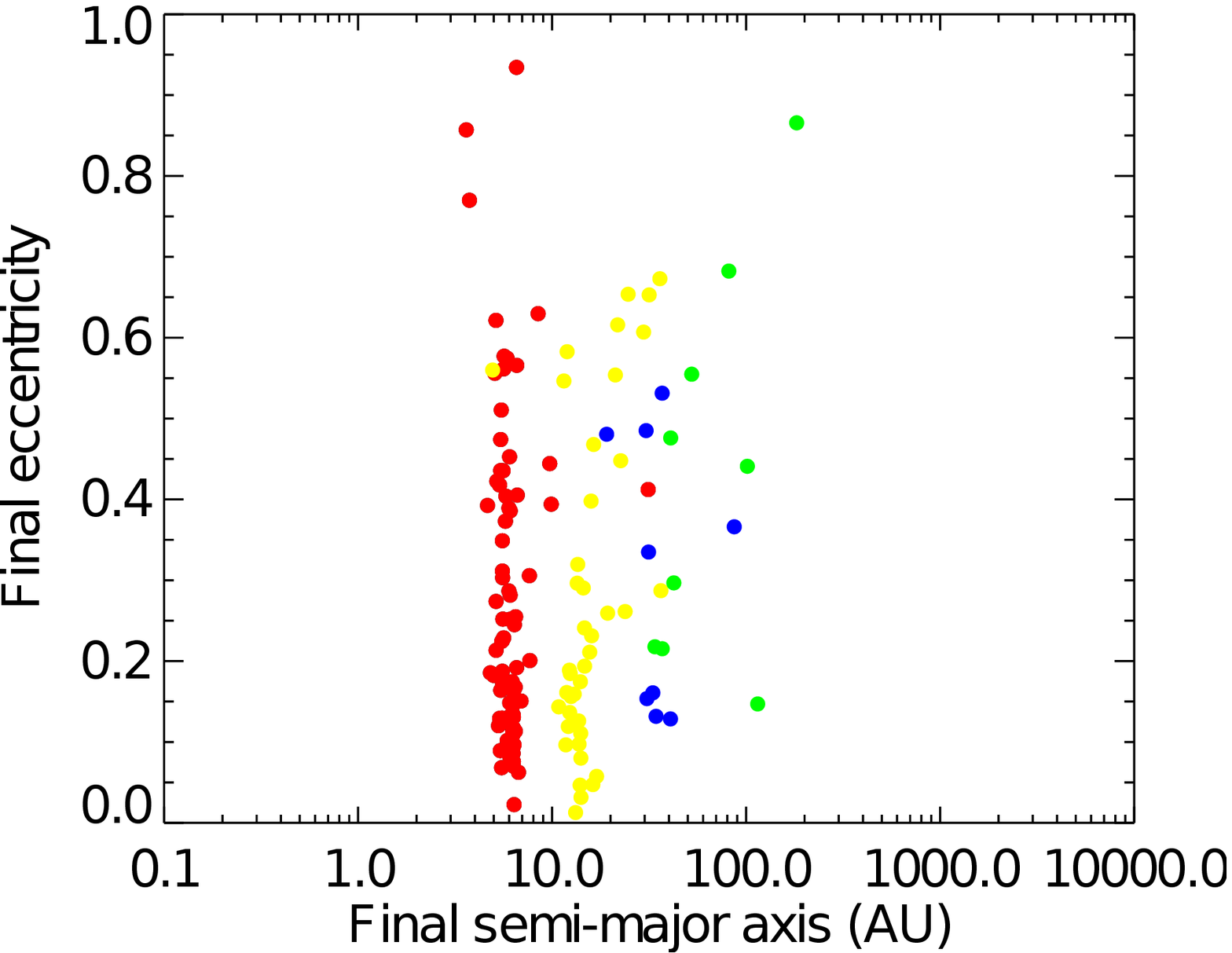}
  \end{tabular}
  \caption{Eccentricity versus semi-major axis of the bound planets at the end of the simulations, for model~1 ({\em top}) and model~2 ({\em bottom}). The different colours in the top panel indicate the initial semi-major axis of 1~AU (red), 2.6~AU (yellow), 6.5~AU (green), 16.6~AU (blue), and 42.3~AU (black). In the bottom panel the red, yellow, blue, and green colours represent Jupiter, Saturn, Uranus, and Neptune. At the end of the simulations, a substantial number of planets has obtained a higher eccentricity and a wider orbit. 
    \label{figure:a_and_e} }
\end{figure}

Figure~\ref{figure:bound_time}
shows the fraction of remaining bound planets decreases smoothly with time. The equal-mass system (model~1) looses roughly one~planet per 100~Myr, while the planet ejection rate is roughly two planets per 100~Myr for model~2. The latter set shows a decrease in the escape rate over time due to the reduced number of planets in the system. At the end of the simulations, each host stars has, on average, less than three planets remaining for model~1, and on average slightly more than one planet for model~2. The downward trend is still present at the end of the simulations, which simply indicates that additional planets will be removed from the system if it were to remain part of a star clusters beyond $\sim 150$~Myr.

Distant stellar encounters perturb each of the planetary orbits to a certain degree. This generally result in a slight increase in the semi-major axis, the eccentricity, and the inclination. The cumulative effect of many encounters can substantially perturb the outer planetary orbits, which can in turn result in strong planet-planet interactions. A number of planets obtain orbits that ultimately result in a physical collision with the central star. Also, some of the planets escape from the planetary systems and become a free-floating planet in the star cluster after perturbations. 

\begin{figure}
  \centering
  \begin{tabular}{c}
  \includegraphics[width=0.45\textwidth,height=!]{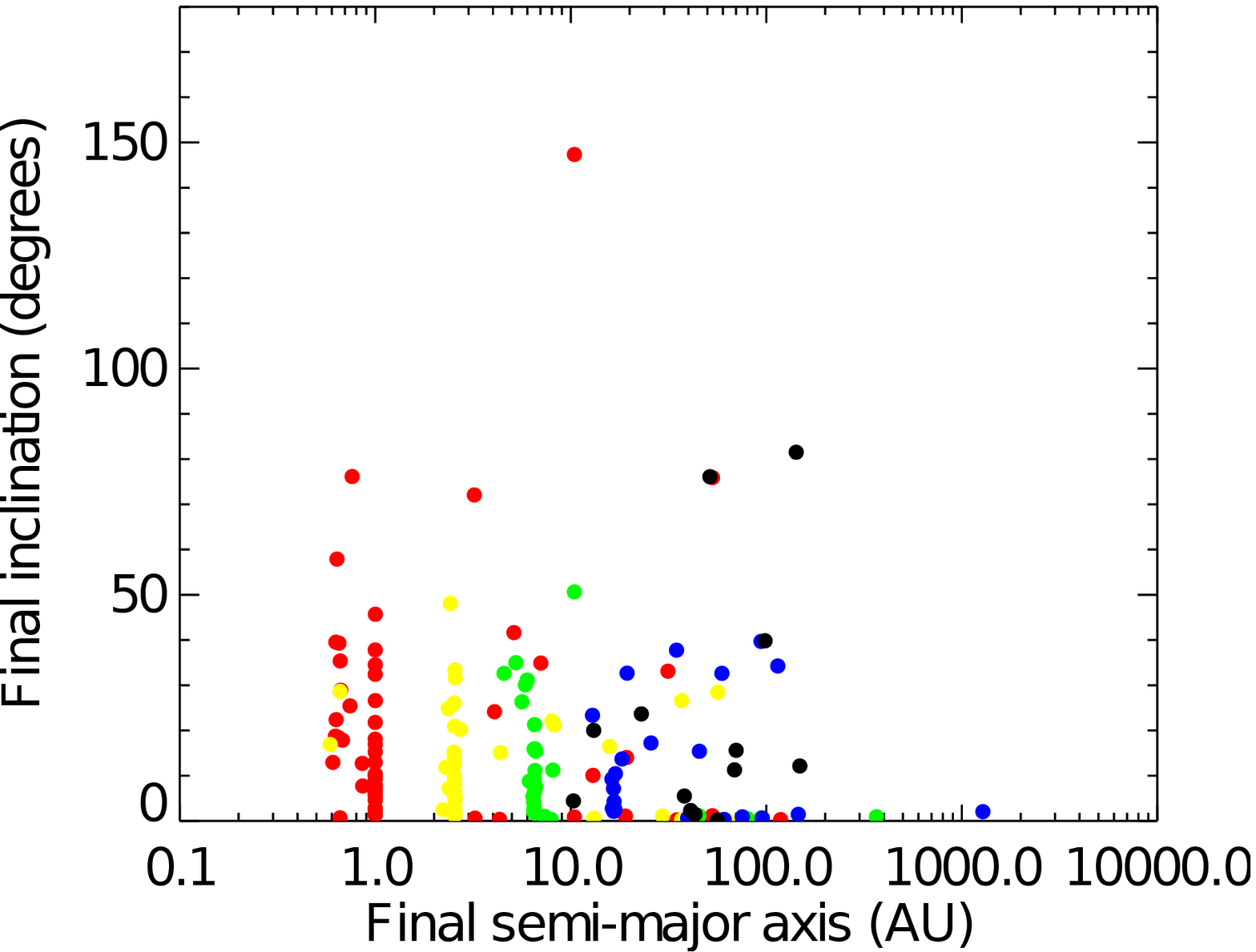} \\
  \includegraphics[width=0.45\textwidth,height=!]{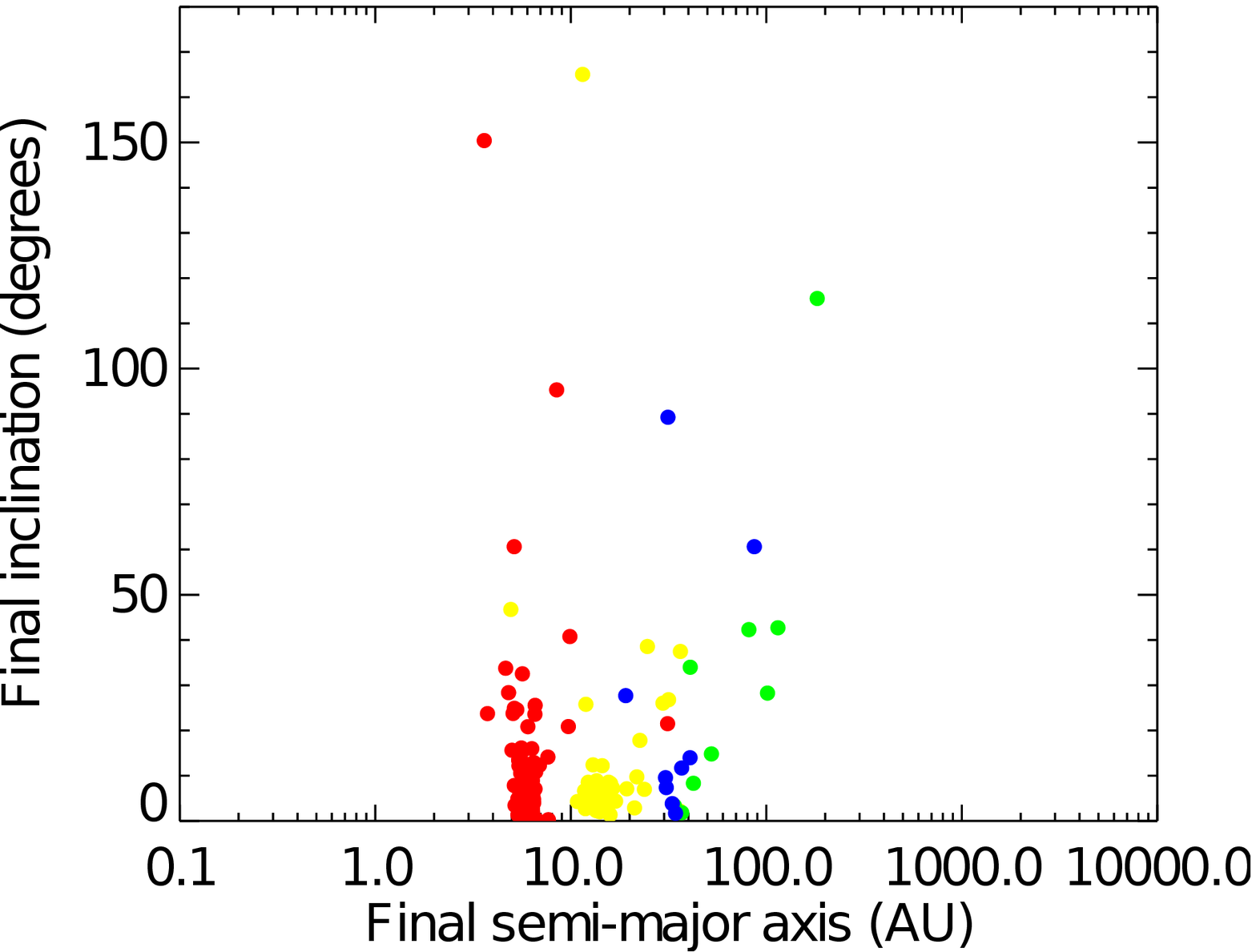}
  \end{tabular}
  \caption{Inclination versus semi-major axis for the bound planets at the end of the simulations, for model~1 ({\em top}) and model~2 ({\em bottom}). Many planets have obtained both a higher inclination due to stellar encounters, although the vast majority still have $i<40^\circ$. Note that some planets have obtained retrograde $(i>90^\circ)$ orbits. 
    \label{figure:a_and_i} }
\end{figure}

Figures~\ref{figure:a_and_e} and~\ref{figure:a_and_i}
show the result of eccentricity and inclination distribution as a function of semi-major
axis after 100~Myr. Note that the planets in model~2 (the four gas giants in the Solar system) lack low-eccentricity and low-inclination orbits, which is a consequence of their non-zero initial conditions. The chimney-like structures illustrate that it is much easier to change the eccentricity and inclination than the semi-major axis. The reason for this is that it is easier for neighbouring planets (and encountering stars) to exchange angular momentum, rather than orbital energy. It is worth mentioning that planets with an inclination larger than 90 degrees have retrograde orbits, and that all planets, including the innermost ones, have a chance of obtaining a retrograde orbit.

\begin{figure}
  \centering
  \begin{tabular}{c}
    \includegraphics[width=0.45\textwidth,height=!]{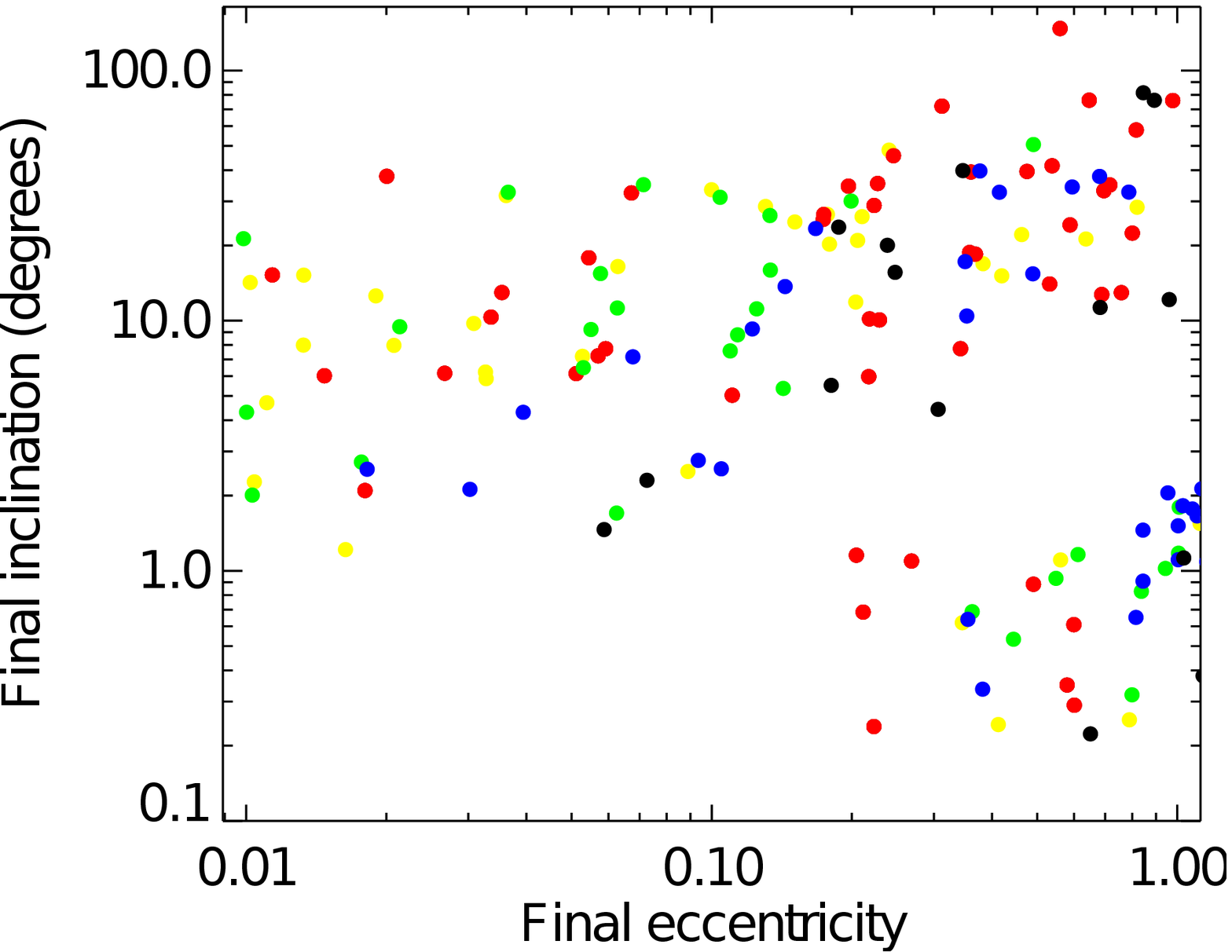} \\
    \includegraphics[width=0.45\textwidth,height=!]{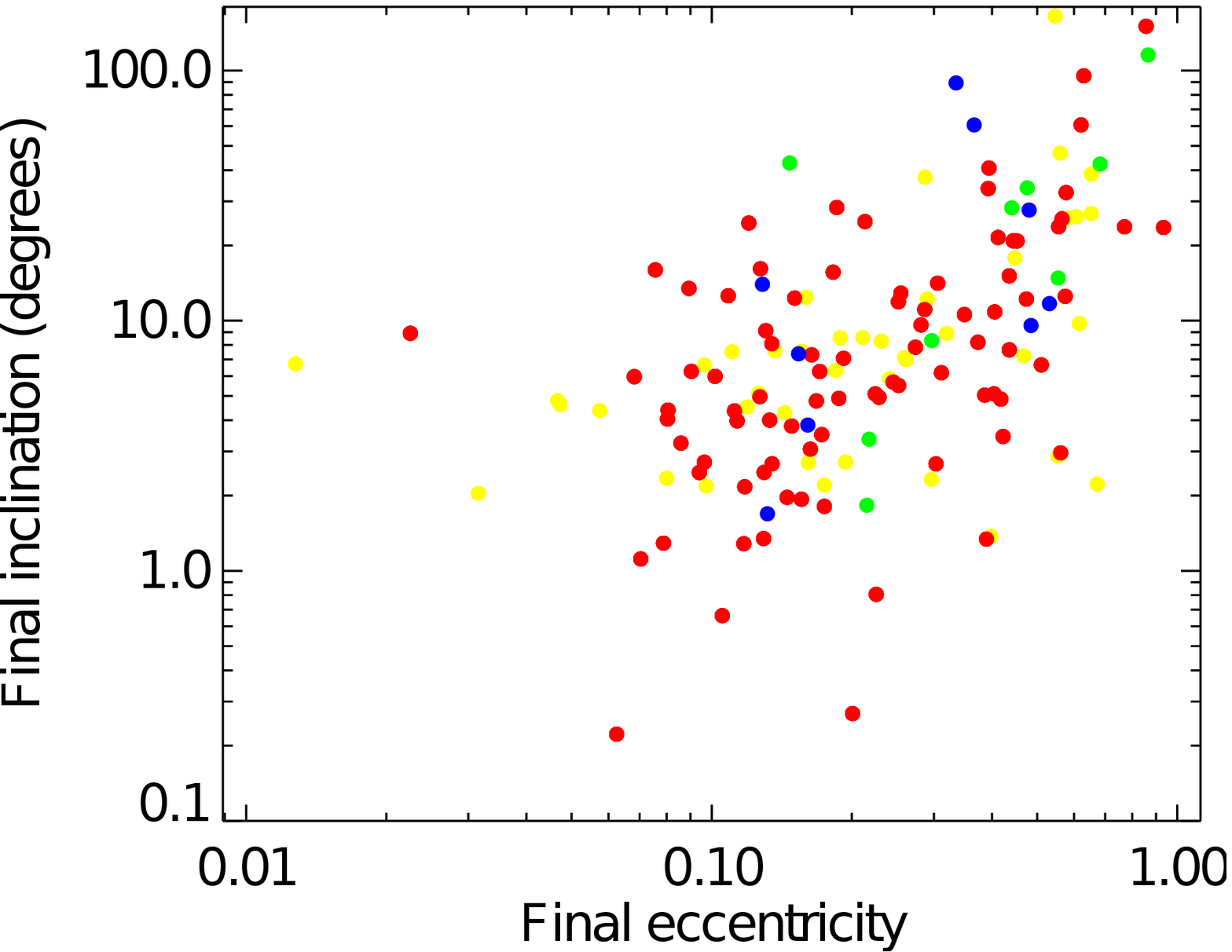}
    \end{tabular}
  \caption{Inclination versus eccentricity of the bound planets at the end of the simulations, for model~1 ({\em top}) and model~2 ({\em bottom}). 
      \label{figure:i_and_e} }
\end{figure}

Figure~\ref{figure:i_and_e} shows the distribution of inclination angles versus eccentricities of all remaining planets in the ensemble of models. We can see that, as expected, perturbed orbits both obtain higher eccentricities and higher inclinations. Note that there is no clear correlation with initial semi-major axis (indicated by the colour of the dots). Almost all highly inclined and retrograde orbits have a large eccentricity, which may ultimately lead to Kozai cycles.

\subsection{Ejected planets}

\begin{figure}
  \centering
  \begin{tabular}{c}
  \includegraphics[width=0.45\textwidth,height=!]{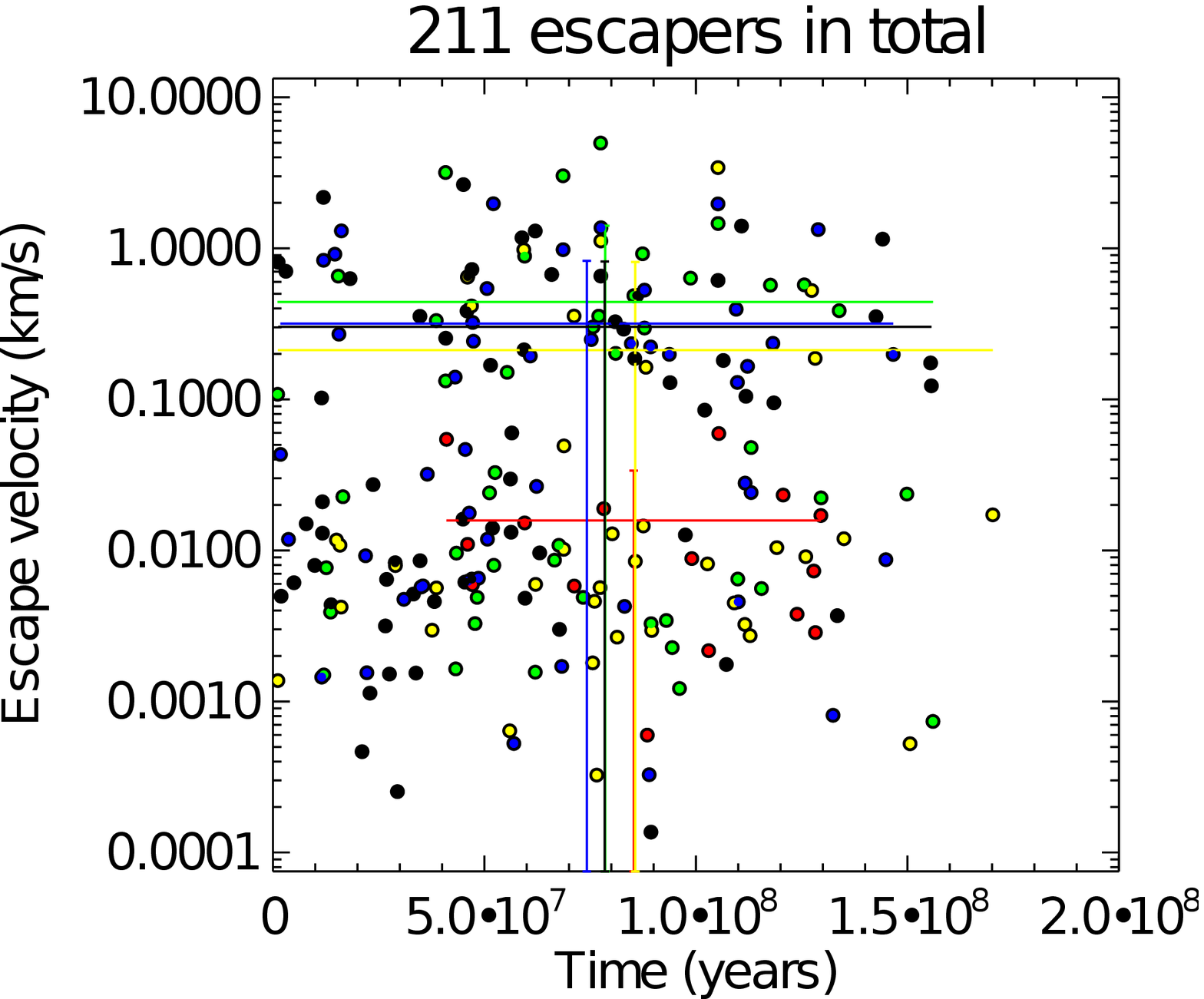} \\
  \includegraphics[width=0.45\textwidth,height=!]{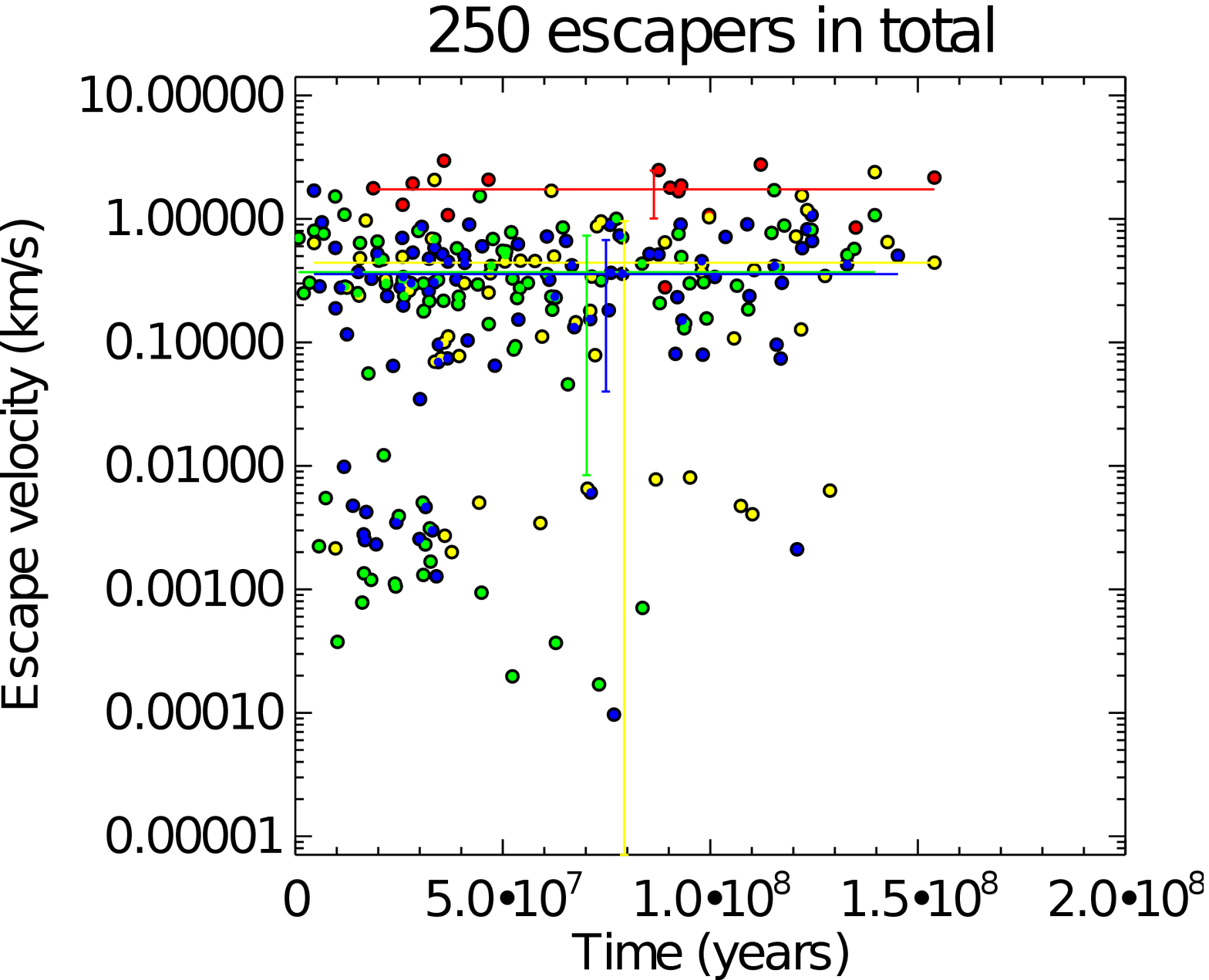}
  \end{tabular}
  \caption{The velocities of the ejected planets as a function of time, for model~1 ({\em top}) and model~2 ({\em bottom}). Most of the inner planets (red data points) have rather low escape velocities, as compared to the high escape velocities of the outer planets (yellow, green, blue and black data points). The crosses indicate for each planet the time interval at which the planets escape and the standard deviation of their escape velocities. 
    \label{figure:escapevelocities} }
\end{figure}

The ejection velocity distribution of the escaping planets is shown in Figure~\ref{figure:escapevelocities}. Note that there is no strong correlation between the ejection velocity and the initial semi-major axis. There is a weak correlation between the initial semi-major axis and the escape velocity. For the stars in model~1, the escaped innermost planets have a smaller escape velocity, which can be attributed to the deeper potential well which they have to escape from. The opposite result is seen for the Jupiters in model~2, which have a larger escape velocity than the other planets. This apparent contradiction can be explained by the mechanism at which these planets escape. In model~1, the innermost planets escape almost exclusively as a result of planet-planet scattering. In model~2, on the other hand, the Jupiters, which are substantially more massive than the other planets, only escape as a result of direct interactions with encountering stars (see also \S~\ref{section:singlemulti}).

We can see that many of the survived planets have relatively large eccentricities and inclinations (Figure~\ref{figure:a_and_e} and Figure~\ref{figure:a_and_i}). These highly eccentric and highly inclined orbits in multi-planet system are unstable and might trigger escapers or collisions in the next several million years once their angular momentum deficit is high enough. 

From the range in time at which the planets escape, indicated with the horizontal lines in Figure~\ref{figure:escapevelocities}, we can see  that the first escapers among the innermost planets in model~1 reach the boundaries of the system only after roughly 50~Myr. The escape velocity distribution among the other four planets is roughly constant in time and independent of their initial semi-major axes. This strongly suggests that escape among the innermost planets is primarily the result of the secular evolution of the planetary system, rather than the direct result of the stellar encounters. For model~2 we observe that each of the planets escape at random times, which indicates that at least a fraction of each of the planets is ejected due to direct interactions with encountering stars.

The planets that are ejected from the system become individual star cluster members. Whether or not these free-floating are immediately ejected from the star cluster as well, depends on their ejection velocity $v_e$, the distance to the star cluster centre $r$. To a lesser extent, it also depends on the direction of the $\vec{v}_e$ and the particle distribution within the cluster. Upon ejection from their planetary system, the planets obtain a velocity $\vec{v} = \vec{v}_e + \vec{v}_s$, where $\vec{v}_s$ is the velocity of the host star in the star cluster. The escape velocity $v_{\rm esc}$ at a distance $r$ from the centre of a star clusters is given by 
\begin{equation}
  v_{\rm esc}(r) = \sqrt{\frac{2GM(r)}{r}} \ ,
\end{equation}
where $G$ is the gravitational constant and $M(r)$ is the enclosed mass within a radius $r$.

When a free-floating planet remains bound to the star cluster, it may be re-captured by another stars \citep{perets2012}, or escape at a later time through ejection or evaporation. Using a microlensing survey, \cite{sumi2011} estimate that free-floating planets are almost twice as abundant as main sequence stars in the Galactic field. \cite{veras2012} find that planet-planet scattering cannot explain the free-floating planet population inferred from microlensing studies, and that other explanations, such as stripping of planets in star clusters, and stellar evolution, may well be necessary to explain the free-floating planet population.

\subsection{Star-planet and planet-planet collisions}

The perturbations induced by the encounters can result in instabilities in the planetary system which may result in star-planet and planet-planet collisions. We list the fraction of planets that experience collisions in Table~\ref{table:multiplanet}. The error $\delta$ on these fractions is $\delta \approx N^{-1}d\sqrt{d^{-1}+N^{-1}}$, where $d$ is the number of ejections, collisions, or survivors, and $N=100$ is the total number of objects in the sample \citep[e.g.,][p.\,108]{kouwenhoven2006}. Here we have made the large-number assumption, and find that the error typically amounts to $\delta = 5-10\%$ for the data listed in the table.

Table~\ref{table:multiplanet} shows that in the equal-mass case (model~1), the inner planets have a substantially larger probability of experiencing a physical collision than the outer planets, which are more frequently ejected from the system. For the innermost planet, the number of escapers ($15\%$) and the number of collisions with the central star ($12\%$) are comparable. The vast majority of the collisions are between planets and the central star, while the collisions between two planets are rarely found in our simulations. The only case of a planet-planet collision is caused by a rare close encounter. Physical collisions are less frequent for model~2, where only 3\% of the Saturns and 3\% of the Neptunes collide with the central star. Collisions between Jupiter and the central star do not occur, as none of the other, lower-mass planets in the system are able to perturb Juptiter's orbit enough.

Among additional tests with low-velocity encountering star in a near parabolic orbit with closest approach within 100~AU, we find that planet-planet collision can occasionally occur. These results demonstrate that planet-planet interactions (rather than direct interactions with encountering stars) play an important role for the long-term evolution of perturbed planetary systems.

\begin{table}
  \begin{tabular}{c cccc}
    \hline
    $a_{\rm initial}$    & Escapers	      & Collisions       & Collisions       & Survivors \\
 	                    	 &      			& w/ star        	& w/ planets	& \\
    \hline 
    1.0~AU               & 15~\%		      & 12~\%        	& 0~\%        	& 73~\% \\
    2.6~AU               & 31~\%		      & 7~\%		&0~\%	        & 62~\% \\
    6.5~AU               & 43~\%             & 5~\%                  & 0~\%		& 52~\% \\
    16.6~AU		 & 51~\%      	      & 2~\%             	& 0~\%		& 47~\% \\
    42.3~AU		& 71~\%		      & 1~\%		& 0~\%	      & 28~\% \\
    \hline
 5.2~AU & 15~\% & 0~\% & 0~\% & 85~\% \\
9.6~AU & 56~\% & 3~\% &0~\% & 41~\% \\
19.2~AU & 91~\% & 0~\% & 0~\% & 9~\% \\
30.1~AU & 88~\% & 3~\% & 0~\% & 9~\% \\
     \hline
  \end{tabular}
  \caption{The outcome of the multi-planet simulations for model~1 (top) and model~2 (bottom). Note that that inner planets have substantially smaller survival chances than in the single-planet case (see Table~\ref{table:singleplanet}).\label{table:multiplanet} }
\end{table}

\begin{table}
  \begin{tabular}{c cccc}
    \hline
    $a_{\rm initial}$    & Escapers	      & Collisions	& Collisions           & Survivors \\
	                     	&               		& w/ star		& w/ planets              & \\
    \hline 
    1.0~AU               & 3~\%		  & 0~\%          & N/A   			& 97~\% \\
    2.6~AU               & 7~\%		  & 0~\%		& N/A    		 	& 93~\% \\
    6.5~AU               & 15~\%        	 & 0~\%             & N/A 			& 85~\% \\
    16.6~AU		 & 46~\%          & 0~\%            & N/A 			& 54~\% \\
    42.3~AU		 & 77~\%	      	& 1~\%	    	& N/A 			& 22~\% \\
    \hline
 5.2~AU & 14~\% & 0~\% & N/A & 86~\% \\
9.6~AU & 24~\% & 0~\% & N/A & 76~\% \\
19.2~AU & 53~\% & 0~\% & N/A & 47~\% \\
30.1~AU & 60~\% & 0~\% & N/A & 40~\% \\	\hline
  \end{tabular}
  \caption{The evolution of single-planet systems in an open cluster environment for model~1 (top) and model~2 (bottom). Note the collisions here exclusively represent mergers between a planet and the central star. \label{table:singleplanet} }
\end{table}

\subsection{Multi-planet systems versus single-planet systems} \label{section:singlemulti}

We study the evolution of multi-planet systems in dense stellar environments. In order to demonstrate the importance of multiplicity and planet-planet scattering in these systems, we carry out additional simulations in which we follow the evolution of single-planet systems in identical environments. In this experiment we carry out simulations in which the planetary orbits are as described in \S~\ref{section:initialconditions}, but in this case only with for stars with a single planet. We use exactly the same properties of the encountering stars and encounter intervals, such that we are able to make a proper comparison.

The results for the evolution of the single-planet systems are listed in Table~\ref{table:singleplanet}. The differences with those of the multi-planet systems 
(Table~\ref{table:multiplanet}) are obvious: planet-planet scattering is of great importance to the evolution of multi-planet systems. Among the close-in planets in model~1 we observe many more escapers and collisions in the multi-planet case, with respect to the single-planet case. The differences are smaller for the outermost planets, as these are mostly affected by direct interactions with encountering stars. Our data hints that outer planets have a slightly higher survival chance in multi-planet systems. One possible explanation for this is the interaction with the inner planetary system, which may exchange angular momentum with a strongly perturbed outer planet, which may mildly reduce its chances of escape. However, statistics should be improved in order to make a clear statement about this difference. In the case of the four Solar system planets (model~2), a similar trend is observed. One important difference is that, although the evolution of the outer three planets is substantially affected by the multiplicity, the Jupiters are unaffected. The reason for this is that the Jupiters are much more massive than the other planets, and therefore relatively unaffected by planet-planet scattering. For this reason, the surviving fraction of Jupiters is roughly $85\%$, irrespective of whether other, lower-mass planets are present in the system.

\section{Conclusions} \label{section:conclusions}

We have studied the evolution of planetary systems in star clusters similar to the ONC over
 an evolutionary time of the order of $10^8$ years. We study two types of planetary systems in particular. In the first set of simulations (model~1) we study the effect of encounters between
 planetary systems consisting of five Jupiter-mass planets in the semi-major axis range of $1-50$~AU and single stars that approach these planetary systems within a distance less than $1000$~AU. In the second set of simulations, we consider systems similar to our own Solar system, containing the Sun, Jupiter, Saturn, Uranus, and Neptune. The interactions
 are integrated using the classical Aarseth-Mikkola chain integrator
 in regularised coordinates with high efficiency and accuracy. 
 For the time interval between two subsequent interactions the MERCURY integration package is used to follow the
 internal secular evolution of the planetary systems. Our main results are as
follows: 
\begin{enumerate}
\item In multi-planet systems in open star clusters, planets of all periods are either directly or indirectly affected by stellar encounters. The inner planetary orbits are affected as a result of interaction with perturbed outer planets.
\item For the equal-mass system (model~1), the planets starting at $a\approx 42$~AU, approximately $25\%$ survives, while $\sim 75\%$ escapes, both in the single-planet and multi-planet cases. For those planets starting out at $a=1$~AU in a multi-planet system, approximately $15\%$ escapes, $12\%$ collides with the central star, and 73\% survives. In the single-planet case, however, $93\%$ of the planets at $a=1$~AU survives, $3\%$ escapes, and we observe no collisions. 
\item For the four giant planets in our Solar system (model~2), we observe a similar trend, although we see that Jupiter is not substantially affected by planet-planet encounters, because of its large mass with respect to the other three planets. 
\item The planetary systems loose their planets at a rate of one per 100~Myr (model~1) and two per 100~Myr (model~2), respectively. A gradual decrease in the escape rate is seen over time, due to the smaller number of remaining planets in each system.
\item Although our results apply to the two specific planetary systems a given star cluster environment, these results do demonstrate that multiplicity cannot be ignored when studying the evolution of planetary systems in crowded environments.
\item Most of the planets' semi-major axes are increased slightly after the
  each encounters. The majority of the planets obtain inclined and eccentric orbits. Generally, these effects are strongest for the outer planets. Typically a few percent of the planets are found in highly inclined or retrograde orbits at the end of the simulations. This applies to planets of any initial semi-major axis, and can be attributed to both the stellar encounters and planet-planet scattering. The latter process is dominant for the inner planets. 
\item The escaping planets are ejected with velocities up to several \kms, but, as Figure~\ref{figure:escapevelocities} shows, the majority of these newly formed free-floating planets are unable to obtain velocities beyond the escape velocity of their host star cluster. In addition, we find that the escape velocity increases with the initial semi-major axis of the planets.
\end{enumerate}
In summary, the presence of multiple planets in a planetary system results in a
  significantly different evolution as compared to single-planet
  systems, as a result of planet-planet interactions. Our results on the orbital evolution and also our planet liberation rates per crossing time are in fair agreement with the earlier direct cluster simulations of \cite{spurzem2009}, who find in their Table~3 approximately one planet liberation per crossing time for their wide planets (up to $50$~AU). 

\section{Discussion} \label{section:discussion}

Our model is a simplification of reality in a number of ways, and our assumptions should be improved and addressed in future work, ideally through directly modelling multi-planet systems in star clusters. 

We draw masses from an invariant Chabrier mass distribution (Eq.~\ref{equation:chabrier}), we draw the relative velocities from a local Maxwellian distribution function, based on the average velocity dispersion of the Plummer model (Eq.~\ref{equation:maxwellian}), we draw impact parameters from a linear probability distribution function (Eq.~\ref{equation:impactparameter}), and we assume that subsequent encounters are uncorrelated, well separated in time (Eq.~\ref{equation:poisson}). 
In a real evolving star cluster all these assumptions are to some degree approximations, 
but the full star cluster simulation is a much larger numerical effort than our work here, and to some extent it is more complex to disentangle the effects of individual encounters in a full simulation, as can be seen from the rather complicated procedures used in \cite{spurzem2009} to identify encounters in $N$-body simulations. It should be noted, however, that mass segregation, core collapse and the finite size of the core of a star cluster will lead to a time variation of the encounter parameters, and to a deviation from the simple distribution functions assumed here. In particular, it is known that in real star clusters the velocity distribution function can be far from a Maxwellian \citep{plummer1911,cohn1980}. 

 
The encounter rate in a star cluster decreases with time as a result of expansion. In addition, a star may escape from a star
cluster within a short time, after which the effect of stellar encounters on their
planetary systems are negligible. We obtain approximate evolutionary tracks using the EMACSS code \citep{alexander2012}, and estimate the stellar encounter rate as a function of time under the approximation that the interactions are dominated by gravitational focusing \citep[see][]{malmberg2011}. As expected, the encounter rate gradually decreases over time, by a factor of two after 30~Myr down to a factor of six after 100~Myr. These estimates demonstrate that we may have overestimated the number of encounters typically experienced by a planetary system. On the other hand, star
clusters may form sub-virial, and star forming regions observed to form with a significant amount of substructure in both
the position and the velocity distribution, which can
 boost the number of close encounters during the early stages of star cluster evolution \citep{allison2009}. Finally, if a star cluster follows the evolutionary track listed above, many of its stars will suffer additional encounters until the cluster is completely dissolved. This shows again the need for full $N$-body simulations, which naturally include all processes described above.
Despite these approximations, we have demonstrated the importance of multiplicity of planetary systems, and we expect that improved future simulations will give qualitatively the same result.


We have also neglected the presence of binary systems, despite their presence in star clusters \citep[e.g.,][]{kouwenhoven2005,kouwenhoven2007} and in the Galactic
field \citep{duquennoy1991}. The effect of binary systems is complex, and depends strongly on the semi-major axis $a$ of the binary and the distance of closest approach $p$ of the centre-of-mass of the binary. Encounters with sufficiently tight binaries ($a \ll p$) have a very similar effect as encountering single stars, since they can, to first order, be considered as single bodies with a larger mass (up to a factor two for equal-mass binary systems). Very wide binaries ($a \gg p$) also have a similar effect as single stars, as a planetary system only feels the nearby component of the binary system. The main difference with the encounters of single stars is that, while the centre-of-mass of a wide binary approaches impact parameter $b$, the periastron distance of the nearest binary component may come closer to the planetary system. Complex situations arise in the case of intermediate-period binaries with $a \approx p$. In this case, both components of the binary system can affect the planetary system, but more importantly, highly destructive three-body interactions can occur. Depending on the properties of the binary population in a star clusters, these types of interactions may or may not have an important effect on the survival of planetary systems. Further work is definitely necessary to study the effect of binary systems. However, from the arguments above one can see that the inclusion of binaries, whether they are tight, intermediate, or wide, will generally cause further destruction of planetary systems. 

Finally, it would be interesting to see how our results change when considering the immense diversity among the observed exoplanet systems. Systems with a different number of planets, planetary masses, and relative separations will result in different outcomes. However, the general outcome of our simulations (planets at any orbital separation are either directly or indirectly affected by distant stellar encounters) is unlikely to change.

Future work will combine our ansatz with full star cluster simulations. The secular evolution of planetary systems between encounters in an $N$-body simulation of a large star cluster can be treated by either an adapted version of MERCURY code or by a generalisation of Lagrange's planetary equations, while encounters will take place as in a real star cluster initialised by the time-dependent environment of a full $N$-body simulation.

\section*{Acknowledgments}

We would like to express our gratitude to the anonymous referee, Pei Huang, Thorsten Naab, Douglas Lin, Sverre Aarseth, and Melvyn Davies for constructive discussions. W.H. was supported by Peking University, the Kavli Institute for Astronomy and Astrophysics and the Max-Planck Institute for Astrophysics. M.B.N.K. was supported by
the Peter and Patricia Gruber Foundation through the PPGF fellowship,
by the Peking University One Hundred Talent Fund (985), and by the
National Natural Science Foundation of China (grants 11010237,
11050110414, 11173004). 
This publication was made possible through the support of a grant from the John Templeton Foundation and National Astronomical Observatories of Chinese Academy of Sciences. The opinions expressed in this publication are those of the author(s) do not necessarily reflect the views of the John Templeton Foundation or National Astronomical Observatories of Chinese Academy of Sciences. The funds from John Templeton Foundation were awarded in a grant to The University of Chicago which also managed the program in conjunction with National Astronomical Observatories, Chinese Academy of Sciences. 
R.S. acknowledges support by the Chinese Academy of Sciences Visiting Professorship for Senior International Scientists, grant number 2009S1-5 (The Silk Road Project).

\end{document}